\documentclass[a4paper]{jpconf}
\usepackage{graphicx}
\usepackage{epsf}
\begin{document}

\title[Hadron Spectroscopy with COMPASS at CERN]{Hadron Spectroscopy with COMPASS at CERN}

\author{Karin Sch\"onning for the COMPASS collaboration}

\address{European Organisation for Nuclear Research (CERN), CH-1211 Geneva 23, Switzerland}
\ead{karin.schonning@cern.ch}
\begin{abstract}

The aim of the COMPASS hadron programme is to study the light-quark hadron spectrum, and in particular, to search for evidence of hybrids and glueballs. COMPASS is a fixed-target experiment at the CERN SPS and features a two-stage spectrometer with high momentum resolution, large acceptance, particle identification and calorimetry. A short pilot run in 2004 resulted in the observation of a spin-exotic state with $J^{PC}=1^{-+}$ consistent with the debated $\pi_1(1600)$. In addition, Coulomb production at low momentum transfer data provide a test of Chiral Perturbation Theory. During 2008 and 2009, a world leading data set was collected with hadron beam which is currently being analysed. The large statistics allows for a thorough decomposition of the data into partial waves. The COMPASS hadron data span over a broad range of channels and shed light on several different aspects of QCD. 

\end{abstract}



The aim of hadron physics is a fundamental understanding of all bound and resonant systems which interact via the strong force. At short distances, perturbation theory can be applied and its predictions have been rigorously and successfully tested. At longer distances, the self-coupling of the gluons makes perturbation theory inapplicable and effective field theories such as Chiral Perturbation Theory ($\chi$PT), or advanced computational tools like Lattice QCD, have to be used instead. 
It is generally believed that the key to understand QCD at low and intermediate momentum transfers lies in the hadron spectrum. Apart from the conventional quark-antiquark ($q\overline{q}$) and quark triplet ($qqq$) states within the simple contituent quark model, QCD also allows for hadronic matter with excited gluonic degrees of freedom, provided they are colour neutral. A \textit{glueball} is a state described entirely in terms of gluonic fields with no constituent quarks, whereas a \textit{hybrid} is a meson with a constituent gluonic excitation. The excited glue in a hybrid contributes to its quantum numbers, which enables states with quantum numbers $J^{PC}$ which are forbidden for simple $q\overline{q}$ pairs to be formed. Such states are called \textit{spin-exotics}. Lattice calculations predict a spin-exotic hybrid with mass within 1-2 $\rm{GeV/c^2}$ \cite{closepage}. Some promising candidates have been found in experiments: the $\pi_1(1400)$ seen by BNL, VES and Chrystal Barrel \cite{pi1400}, the $\pi_1(1600)$ reported by BNL and VES \cite{pi1600} and the $\pi_1(2000)$ observed by BNL \cite{pi2000}, but the resonance character of these states is disputed to this day. One goal of the COMPASS hadron programme is to bring clarity into this issue.

The COMPASS (COmmom Muon and Proton Apparatus for Structure and Spectroscopy) spectrometer is located at the M2 beam line from the SPS accelerator at CERN. The physics studied by the muon programme is presented elsewhere in these proceedings \cite{silva}. The basic features of the spectrometer are described in Ref. \cite{compassnim} and the specifics of the hadron programme will be outlined in a forthcoming paper \cite{compasshad}. 

\begin{figure}[ht]
\centering
\includegraphics[width=.8\textwidth]{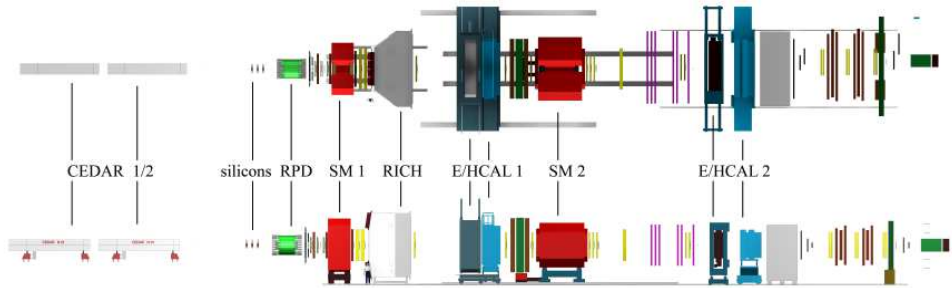}
\vspace{-3mm}
\caption[The COMPASS spectrometer]{Top- and sideview of the COMPASS hadron setup.}
  \label{fig:spectro}
\end{figure}

The data for the hadron programme were collected during 2008/09 at the M2 beam line at the SPS at CERN. The negative hadron beam consisted of 96.8\% $\pi^-$ and 2.4\% $K^-$, whereas the positive consisted of 74.6\% $p$, 24.0\% $\pi^+$ and 1.4\% $K^+$, all with momentum 190 GeV/c. The beam particles were identified by differential Cerenkov detectors (CEDARs) located upstream of the target. The major part of the hadron data were collected using a 40 cm long liquid hydrogen target, but there were also runs with various thin nuclear target discs such as lead, nickel and tungsten. The target was surrounded by a Recoil Proton Detector (RPD) where the recoil target proton was detected. The forward going particles were detected in the two-stage magnetic spectrometer providing precise tracking, RICH particle identification and calorimetry. The layout of the COMPASS spectrometer is shown in Fig. \ref{fig:spectro}.
\vspace{-4mm}
\vspace{-3mm}

At the COMPASS beam energy, 190 GeV/c, three production mechanisms are accessible. In \textit{diffractive dissociation}, reggeon exchange between the target and the beam hadron excites the beam to an intermediate state $X$, which then decays. This is a likely production mechanism for spin exotic hybrids, provided they exist. In \textit{central production}, double reggeon exchange form a state $X$. Double Pomeron exchange is a special case of this process and provides a glue-rich environment which should be suitable for glueball production. In \textit{Coulomb production}, the target radiates a photon which excites the beam hadron. This process is important at low momentum transfer and is a testing ground for $\chi$PT. 


Interesting physics results from the COMPASS hadron programme were obtained already from the three-day pilot run in 2004, using a 190 $GeV/c$ $\pi^-$ beam impinging on a lead target. In the analysis, presented in Ref. \cite{exotic}, diffractive dissociation of the beam pion into an intermediate state $X^-$ was considered, where $X^-$ decays subsequently into a $\pi^-\pi^+\pi^-$ final state. The selected momentum transfer range was $0.1 < t' < 1.0 \,(\rm{GeV/c})^2$, where $t' = t - t_{\rm{min}}$ and $t = (p_{\rm{beam}}-p_X)^{2}$. In this region, the nucleons in the lead nucleus act as quasi-free particles. The data sample comprised $420\,000$ events. The quantum numbers of $X^-$, \textit{i.e.} the spin $J$, parity $P$, C-parity $C$, spin projection $M$ and reflectivity $\epsilon$, were disentangled by Partial-Wave Analysis (PWA). COMPASS uses two independent programs for this task, one developed at Illinois, JINR Dubna and IHEP Protvino \cite{illinois}, and one at Brookhaven \cite{brookhaven}. They have been adapted for COMPASS \cite{rootpwa}. The analysis was performed in two steps: fit in bins of the three-pion invariant mass $m_X$, followed by a mass-dependent fit (for details, see \cite{exotic}). A spin exotic $J^{PC} = 1^{-+}$ wave was observed, consistent with the disputed $\pi_1(1600)$ \cite{pi1600}. A preliminary mass-independent PWA of $\approx$25\% of the available the 2008/09 data set, collected using a hydrogen target and comprising 23 million $\pi^-\pi^+\pi^-$ events, confirms the enhancement in the intensity around $M_X = 1.7\, \rm{GeV/c^2}$ and the phase motion with respect to the $1^{++}$ resonance \cite{haas}. Mass-dependent fitting, background studies of {e.g.} the Deck effect \cite{Deck} and leakage studies are ongoing for more definite conclusions.
 
A striking observation in the $\pi^-\pi^+\pi^-$ data is that the intensity of waves with a given spin projection $M$, depends on the target material. Data collected with lead target in 2009 were compared to hydrogen target data from 2008/09. It was found that for hydrogen, the $M = 1$ states, including the exotic $1^{-+}1^+$, are suppressed with respect to lead data, whereas $M = 0$ are more populated in hydrogen, giving a sum of the $M$ substates which remains unchanged \cite{haas}. 

\begin{figure}[htb]
 \centering
\includegraphics[width=.25\textwidth]{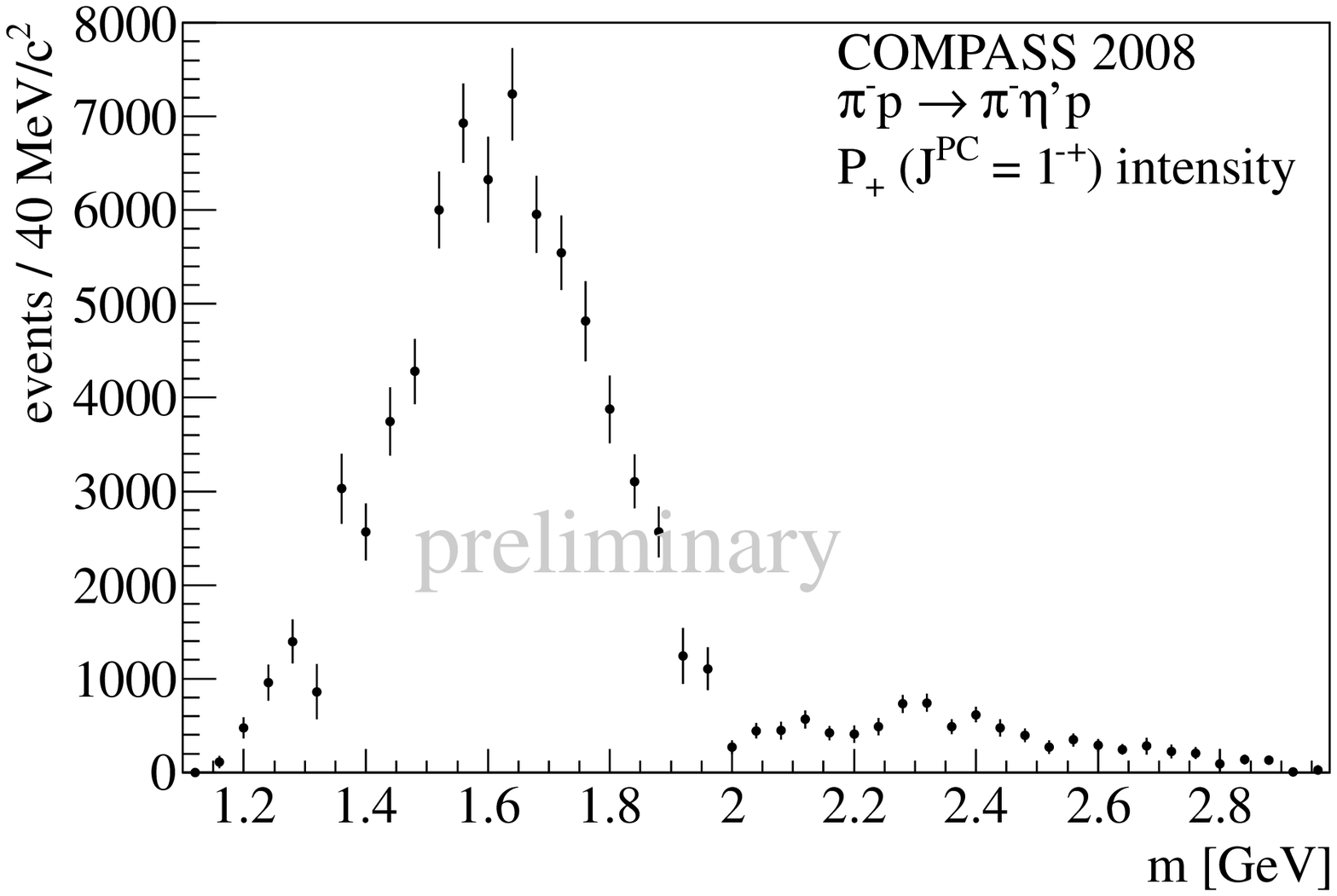}\includegraphics[width=.25\textwidth]{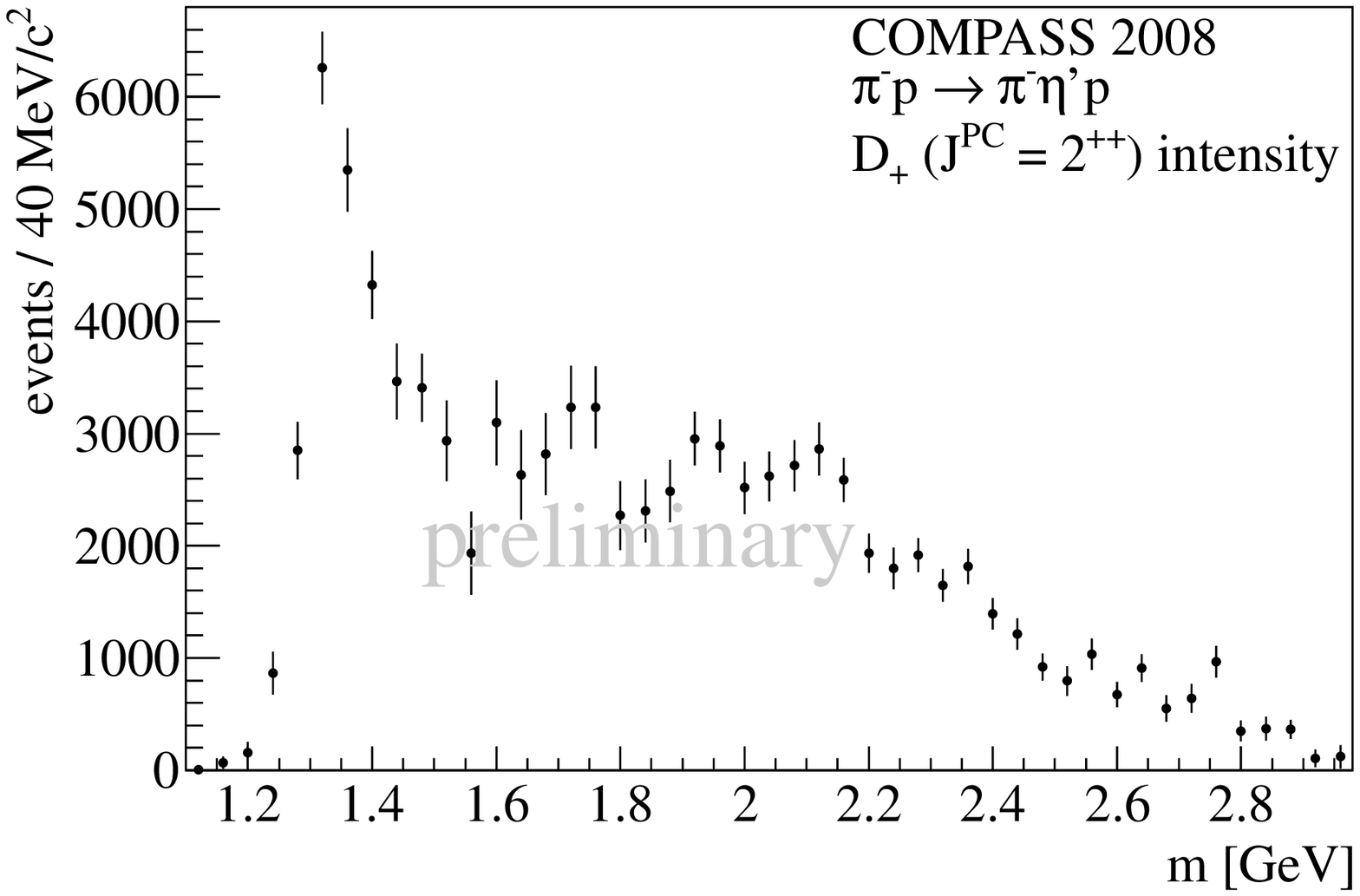}\includegraphics[width=.25\textwidth]{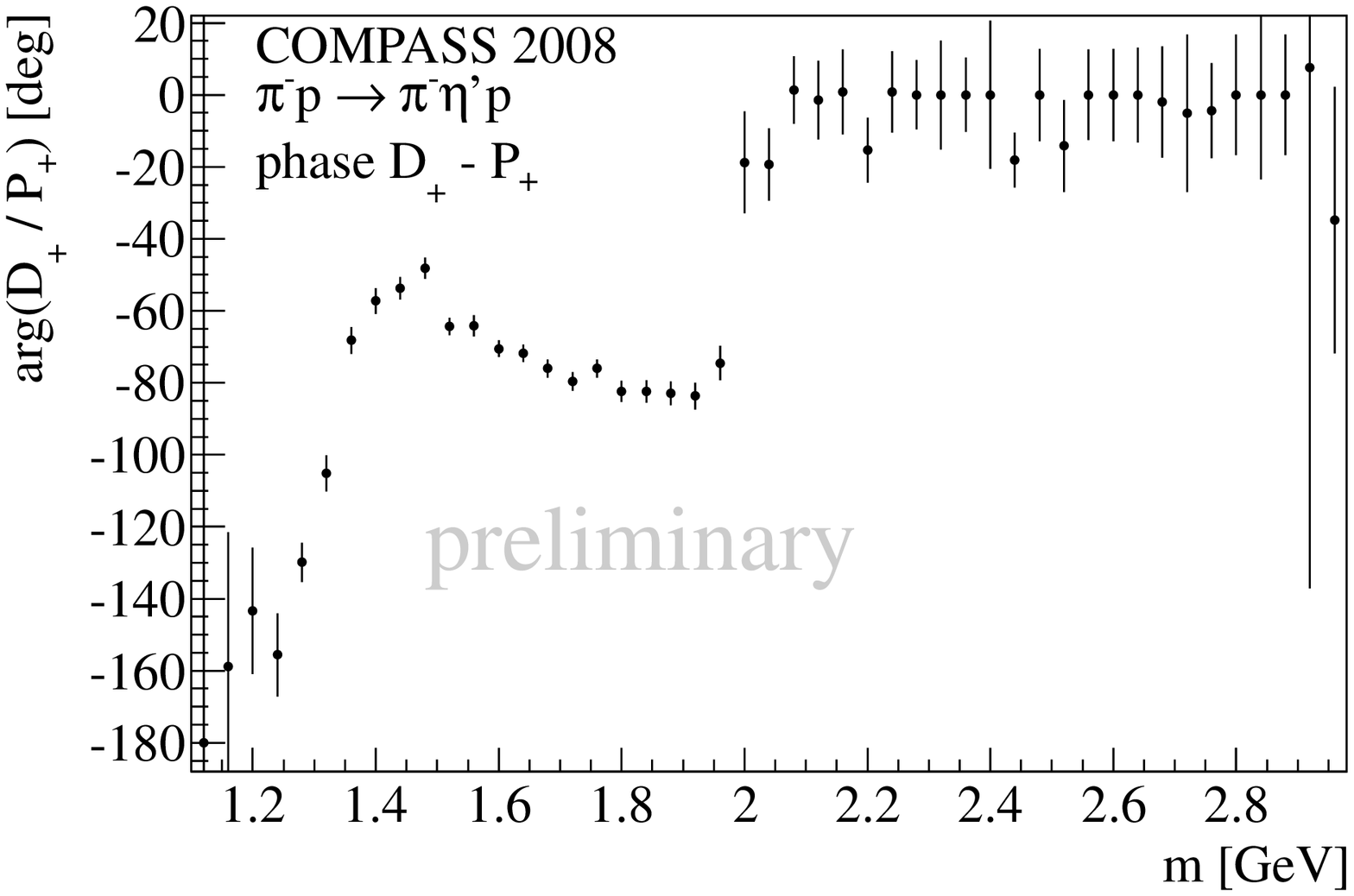}\includegraphics[width=.25\textwidth]{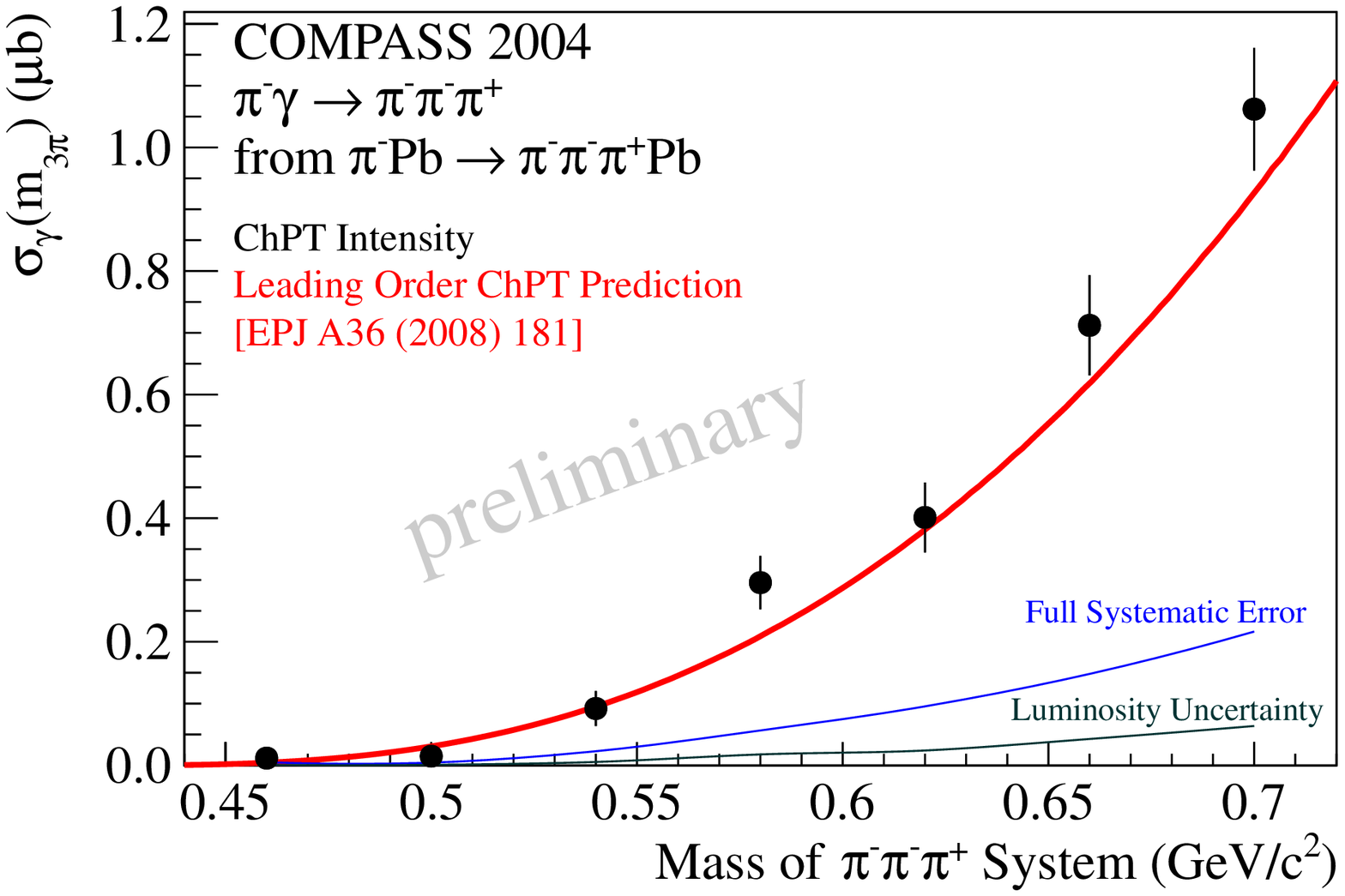}
\vspace{-7mm}
\hspace{-5cm}
\caption{First three from the left show the $\eta'\pi^{-}$ data \cite{schlueter}: The intensity of the $1^{-+}$ and the intenisty of the $2^{++}$ wave, respectively, and the phase difference between $2^{++}$ and $1^{-+}$. Right: The $\gamma\pi^- \rightarrow \pi^-\pi^+\pi^-$ cross section \cite{steffi}. The points are COMPASS 2004 data, the red curve $\chi$PT predictions \cite{friedrich}, the blue the full systematic uncertainty and the back curve the uncertainty from luminosity.}
  \label{fig:chiral}
\end{figure}

The two electromagnetic calorimeters allow for studies of neutral final states. One example is the $\pi^0\pi^0\pi^-$ final state, which provides an important consistency check of the results in the $\pi^-\pi^+\pi^-$ channel. Preliminary PWA's, where the  $\pi^-\pi^+\pi^-$ and the $\pi^0\pi^0\pi^-$ final states are compared, show good agreement between the observed wave intensities and the predictions using isospin- and Bose symmetry \cite{nerling2pi0}. Another channel with neutral particles in the final state is the $\eta'\pi^-$ with $\eta'$ decaying into $\eta\pi^+\pi^-$. The first PWA of these data show a strong $1^{-+}$ wave, shown in Fig. \ref{fig:chiral} where also the intensity of the $2^{++}$ wave and their phase difference are given. However, further studies are needed in order to draw conclusions about the resonance interpretation of the $1^{-+}$ \cite{schlueter}. COMPASS can also confirm the decay of $a_4(2040)$ into  $\eta'\pi^-$ observed by BNL \cite{a2040}. 

The possibility to tag beam kaons with the CEDARs in combination with the RICH identification of final state kaons makes COMPASS an excellent tool for studying kaon diffraction. Recent results from the ongoing PWA of the $K^-\pi^+\pi^-$ final state \cite{jasinski} show a spectrum of states which is mostly in agreement with previous results from the ACCMOR collaboration \cite{wa03}. Channels with kaons in the final state, \textit{e.g.} $\pi^- p \rightarrow p(K\overline{K}\pi)\pi^-$, are also being studied \cite{bernhardnerling}. COMPASS can provide about an order of magnitude more events and a cleaner sample than previous measurement by BNL. This opens up possibilities to study \textit{e.g} the $f_1(1420)\pi^-$ system for the first time. 

The cross section of pion production in $\gamma\pi^- \rightarrow \pi^-\pi^+\pi^-$, a subprocess of $\pi^- Pb \rightarrow \pi^-\pi^+\pi^- Pb$, at low momentum transfer $t' < 0.001 \,(\rm{GeV/c})^2$ has been measured using data from 2004 \cite{steffi}, and the preliminary results, shown to the right in Fig. \ref{fig:chiral}, are in agreement with Leading Order $\chi$PT predictions \cite{friedrich}.

Data collected with the proton beam have been used to measure the ratio between the cross sections of $pp \rightarrow pp \phi$ and $pp \rightarrow pp \omega$. This provides a test at high energy of the Okubo-Iizuka-Zweig (OZI) rule \cite{OZI}. The preliminary result violates the prediction from Ref. \cite{lipkin} by a factor of $\approx$3 \cite{johannes}. The proton beam data are also being used to study baryon spectroscopy \cite{alex} and central production \cite{johannescentral}.

To summarise, the COMPASS hadron programme provides excellent opportunities to study different aspects of QCD. A rich variety of channels are being studied, we can provide more than ten times the world statistics and interesting results have started to emerge.
 
\vspace{-7mm}
\hspace{-12mm}



\end{document}